\pdfoutput=1
\documentclass[12pt,a4paper]{article}

\def\BPV {\ensuremath{B \to PV}\xspace}

\def\sperp {\ensuremath{s_\bot}\xspace}

\def\pzero {\ensuremath{p_0}\xspace}
\def\pone {\ensuremath{p_1}\xspace}
\def\ptwo {\ensuremath{p_2}\xspace}

\def\ACPphi {\ensuremath{\acp(\phiI K^\pm)}\xspace}

\def\pipipi {\ensuremath{{\Bpm \to \pipm \pip \pim}}\xspace}
\def\kpipi {\ensuremath{{\Bpm \to \Kpm \pip \pim}}\xspace}

\def\kkpi {\ensuremath{{\Bpm \to \pipm \Kp \Km }}\xspace}
\def\kkk {\ensuremath{{\Bpm \to \Kpm \Kp \Km}}\xspace}

\def\rhoI   {\ensuremath{\rho(770)^0}\xspace}

\def\kstar  {\ensuremath{\KorKbar^*(892)^0}\xspace}

\def\phiI  {\ensuremath{\phi(1020)}\xspace}

\def\omegaI {\ensuremath{\omega(782)}\xspace}

\def\acp {\ensuremath{A_{\CP}}\xspace}
\usepackage{ifthen} 
\newboolean{pdflatex}
\setboolean{pdflatex}{true} 

\newboolean{articletitles}
\setboolean{articletitles}{true} 

\newboolean{uprightparticles}
\setboolean{uprightparticles}{false} 

\def\paperauthors{LHCb collaboration} 
\def\paperasciititle{Search for direct CP violation in charged charmless B decays to vector mesons} 
\def\papertitle{Search for direct $C\!P$ violation in charged charmless $B \to PV$ decays} 
\def\paperkeywords{{High Energy Physics}, {LHCb}} 
\def\papercopyright{\the\year\ CERN for the benefit of the LHCb collaboration} 
\def\paperlicence{CC BY 4.0 licence}
\def\paperlicenceurl{https://creativecommons.org/licenses/by/4.0/}

\usepackage[top=1in, bottom=1.25in, left=1in, right=1in]{geometry}

\usepackage{microtype}
\usepackage{lineno}  
\usepackage{xspace} 
\usepackage{caption} 

\usepackage{graphicx}  
\usepackage{color}
\usepackage{colortbl}
\graphicspath{{./figs/}} 

\usepackage{amsmath} 
\usepackage{amssymb}
\usepackage{amsfonts}
\usepackage{upgreek} 

\newcommand*\patchAmsMathEnvironmentForLineno[1]{%
\expandafter\let\csname old#1\expandafter\endcsname\csname #1\endcsname
\expandafter\let\csname oldend#1\expandafter\endcsname\csname
end#1\endcsname
 \renewenvironment{#1}%
   {\linenomath\csname old#1\endcsname}%
   {\csname oldend#1\endcsname\endlinenomath}%
}
\newcommand*\patchBothAmsMathEnvironmentsForLineno[1]{%
  \patchAmsMathEnvironmentForLineno{#1}%
  \patchAmsMathEnvironmentForLineno{#1*}%
}
\AtBeginDocument{%
\patchBothAmsMathEnvironmentsForLineno{equation}%
\patchBothAmsMathEnvironmentsForLineno{align}%
\patchBothAmsMathEnvironmentsForLineno{flalign}%
\patchBothAmsMathEnvironmentsForLineno{alignat}%
\patchBothAmsMathEnvironmentsForLineno{gather}%
\patchBothAmsMathEnvironmentsForLineno{multline}%
\patchBothAmsMathEnvironmentsForLineno{eqnarray}%
}

\usepackage{hyperxmp}

\usepackage[pdftex,
            pdfauthor={\paperauthors},
            pdftitle={\paperasciititle},
            pdfkeywords={\paperkeywords},
            pdfcopyright={Copyright (C) \papercopyright},
            pdflicenseurl={\paperlicenceurl}]{hyperref}
\usepackage[colorinlistoftodos,textsize=scriptsize]{todonotes}

\usepackage[bottom,flushmargin,hang,multiple]{footmisc}

\usepackage[all]{hypcap} 

\usepackage{xspace} 
\usepackage{upgreek}

\def\lhcb   {\mbox{LHCb}\xspace}

\def\babar  {\mbox{BaBar}\xspace}
\def\belle  {\mbox{Belle}\xspace}

\def\MagUp {\mbox{\em Mag\kern -0.05em Up}\xspace}

\ifthenelse{\boolean{uprightparticles}}%
{

 \def\Ppi         {\ensuremath{\uppi}\xspace}

 \def\Pchi        {\ensuremath{\upchi}\xspace}                 
 \def\Ppsi        {\ensuremath{\uppsi}\xspace}

 \def\PDelta      {\ensuremath{\Delta}\xspace}                 
 \def\PXi         {\ensuremath{\Xi}\xspace}                 
 \def\PLambda     {\ensuremath{\Lambda}\xspace}                 
 \def\PSigma      {\ensuremath{\Sigma}\xspace}                 
 \def\POmega      {\ensuremath{\Omega}\xspace}                 
 \def\PUpsilon    {\ensuremath{\Upsilon}\xspace}

 \def\PB      {\ensuremath{\mathrm{B}}\xspace}                 
                  
 \def\PD      {\ensuremath{\mathrm{D}}\xspace}

 \def\PJ      {\ensuremath{\mathrm{J}}\xspace}                 
 \def\PK      {\ensuremath{\mathrm{K}}\xspace}

 \def\Pc      {\ensuremath{\mathrm{c}}\xspace}

 \def\Pi      {\ensuremath{\mathrm{i}}\xspace}

 \def\Ps      {\ensuremath{\mathrm{s}}\xspace}

 \def\thebaroffset{0.0em}
}
{

 \def\Ppi         {\ensuremath{\pi}\xspace}

 \def\Pchi        {\ensuremath{\chi}\xspace}                 
 \def\Ppsi        {\ensuremath{\psi}\xspace}                 
                  
 \mathchardef\PDelta="7101
 \mathchardef\PXi="7104
 \mathchardef\PLambda="7103
 \mathchardef\PSigma="7106
 \mathchardef\POmega="710A
 \mathchardef\PUpsilon="7107
                  
 \def\PB      {\ensuremath{B}\xspace}                 
                  
 \def\PD      {\ensuremath{D}\xspace}

 \def\PJ      {\ensuremath{J}\xspace}                 
 \def\PK      {\ensuremath{K}\xspace}

 \def\Pc      {\ensuremath{c}\xspace}

 \def\Pi      {\ensuremath{i}\xspace}

 \def\Ps      {\ensuremath{s}\xspace}

 \def\thebaroffset{0.18em}
}
\newcommand{\offsetoverline}[2][\thebaroffset]{\kern #1\overline{\kern -#1 #2}}%

\makeatletter
\ifcase \@ptsize \relax
  \newcommand{\miniscule}{\@setfontsize\miniscule{4}{5}}
\or
  \newcommand{\miniscule}{\@setfontsize\miniscule{5}{6}}
\or
  \newcommand{\miniscule}{\@setfontsize\miniscule{5}{6}}
\fi
\makeatother

\DeclareRobustCommand{\optbar}[1]{\shortstack{{\miniscule (\rule[.5ex]{1.25em}{.18mm})}
  \\ [-.7ex] $#1$}}

\def\squark    {{\ensuremath{\Ps}}\xspace}

\def\cquark    {{\ensuremath{\Pc}}\xspace}

\def\pion   {{\ensuremath{\Ppi}}\xspace}

\def\pip    {{\ensuremath{\pion^+}}\xspace}
\def\pim    {{\ensuremath{\pion^-}}\xspace}
\def\pipm   {{\ensuremath{\pion^\pm}}\xspace}

\def\kaon    {{\ensuremath{\PK}}\xspace}

\def\KorKbar {\kern \thebaroffset\optbar{\kern -\thebaroffset \PK}{}\xspace}

\def\Kp      {{\ensuremath{\kaon^+}}\xspace}
\def\Km      {{\ensuremath{\kaon^-}}\xspace}
\def\Kpm     {{\ensuremath{\kaon^\pm}}\xspace}

\def\D       {{\ensuremath{\PD}}\xspace}

\def\DorDbar {\kern \thebaroffset\optbar{\kern -\thebaroffset \PD}\xspace}

\def\Dp      {{\ensuremath{\D^+}}\xspace}
\def\Dm      {{\ensuremath{\D^-}}\xspace}

\def\DpDm    {\ensuremath{\Dp {\kern -0.16em \Dm}}\xspace}

\def\B       {{\ensuremath{\PB}}\xspace}

\def\BorBbar {\kern \thebaroffset\optbar{\kern -\thebaroffset \PB}\xspace}

\def\Bd      {{\ensuremath{\B^0}}\xspace}

\def\BdorBdbar {\kern \thebaroffset\optbar{\kern -\thebaroffset \Bd}\xspace}
\def\Bu      {{\ensuremath{\B^+}}\xspace}
\def\Bub     {{\ensuremath{\B^-}}\xspace}
\def\Bp      {{\ensuremath{\Bu}}\xspace}
\def\Bm      {{\ensuremath{\Bub}}\xspace}
\def\Bpm     {{\ensuremath{\B^\pm}}\xspace}

\def\Bs      {{\ensuremath{\B^0_\squark}}\xspace}

\def\BsorBsbar {\kern \thebaroffset\optbar{\kern -\thebaroffset \Bs}\xspace}

\def\jpsi     {{\ensuremath{{\PJ\mskip -3mu/\mskip -2mu\Ppsi}}}\xspace}

\def\chiczero {{\ensuremath{\Pchi_{\cquark 0}}}\xspace}

\def\Y#1S{\ensuremath{\PUpsilon{(#1S)}}\xspace}

\def\LorLbar     {\kern \thebaroffset\optbar{\kern -\thebaroffset \PLambda}\xspace}

\def\to                 {\ensuremath{\rightarrow}\xspace}

\def\CP                {{\ensuremath{C\!P}}\xspace}
\def\CPT               {{\ensuremath{C\!PT}}\xspace}

\def\AT#1     {\ensuremath{A_{\mathrm{T}}^{#1}}\xspace}           

\def\C#1      {\ensuremath{\mathcal{C}_{#1}}\xspace}                       
\def\Cp#1     {\ensuremath{\mathcal{C}_{#1}^{'}}\xspace}                    
\def\Ceff#1   {\ensuremath{\mathcal{C}_{#1}^{\mathrm{(eff)}}}\xspace}        
\def\Cpeff#1  {\ensuremath{\mathcal{C}_{#1}^{'\mathrm{(eff)}}}\xspace}       
\def\Ope#1    {\ensuremath{\mathcal{O}_{#1}}\xspace}                       
\def\Opep#1   {\ensuremath{\mathcal{O}_{#1}^{'}}\xspace}                    

\newcommand{\aunit}[1]{\ensuremath{\text{\,#1}}}       

\newcommand{\tev}{\aunit{Te\kern -0.1em V}\xspace}
\newcommand{\gev}{\aunit{Ge\kern -0.1em V}\xspace}
\newcommand{\mev}{\aunit{Me\kern -0.1em V}\xspace}
\newcommand{\kev}{\aunit{ke\kern -0.1em V}\xspace}
\newcommand{\ev}{\aunit{e\kern -0.1em V}\xspace}
 
\newcommand{\mevc}{\ensuremath{\aunit{Me\kern -0.1em V\!/}c}\xspace}
\newcommand{\gevc}{\ensuremath{\aunit{Ge\kern -0.1em V\!/}c}\xspace}
\newcommand{\mevcc}{\ensuremath{\aunit{Me\kern -0.1em V\!/}c^2}\xspace}
\newcommand{\gevcc}{\ensuremath{\aunit{Ge\kern -0.1em V\!/}c^2}\xspace}
\newcommand{\gevgevcccc}{\ensuremath{\gev^2\!/c^4}\xspace} 

\def\fb   {\ensuremath{\aunit{fb}}\xspace}
\def\invfb   {\ensuremath{\fb^{-1}}\xspace}

\newcommand{\chisqndf}{\ensuremath{\chi^2/\mathrm{ndf}}\xspace}

\def\gsim{{~\raise.15em\hbox{$>$}\kern-.85em
          \lower.35em\hbox{$\sim$}~}\xspace}
\def\lsim{{~\raise.15em\hbox{$<$}\kern-.85em
          \lower.35em\hbox{$\sim$}~}\xspace}

\def\evtgen     {\mbox{\textsc{EvtGen}}\xspace}

\def\geant      {\mbox{\textsc{Geant4}}\xspace}

\def\photos     {\mbox{\textsc{Photos}}\xspace}

\def\pythia     {\mbox{\textsc{Pythia}}\xspace}

\def\tell1  {TELL1\xspace}
\def\ukl1   {UKL1\xspace}

\newcommand{\lhcborcid}[1]{\href{https://orcid.org/#1}{\hspace*{0.1em}\raisebox{-0.45ex}{\includegraphics[width=1em]{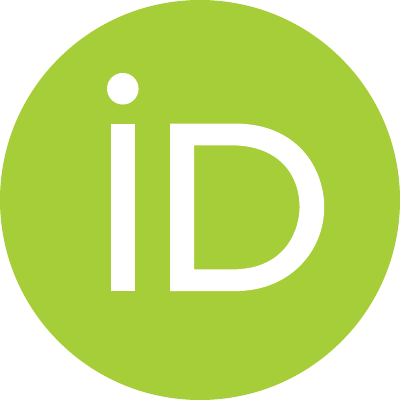}}}}

\usepackage{cite} 
\usepackage{mciteplus}

\usepackage{overpic}

\usepackage{longtable} 

\usepackage{booktabs} 

\begin{document}

\renewcommand{\thefootnote}{\fnsymbol{footnote}}
\setcounter{footnote}{1}


\begin{titlepage}
\pagenumbering{roman}

\vspace*{-1.5cm}
\centerline{\large EUROPEAN ORGANIZATION FOR NUCLEAR RESEARCH (CERN)}
\vspace*{1.5cm}
\noindent
\begin{tabular*}{\linewidth}{lc@{\extracolsep{\fill}}r@{\extracolsep{0pt}}}
\ifthenelse{\boolean{pdflatex}}
{\vspace*{-1.5cm}\mbox{\!\!\!\includegraphics[width=.14\textwidth]{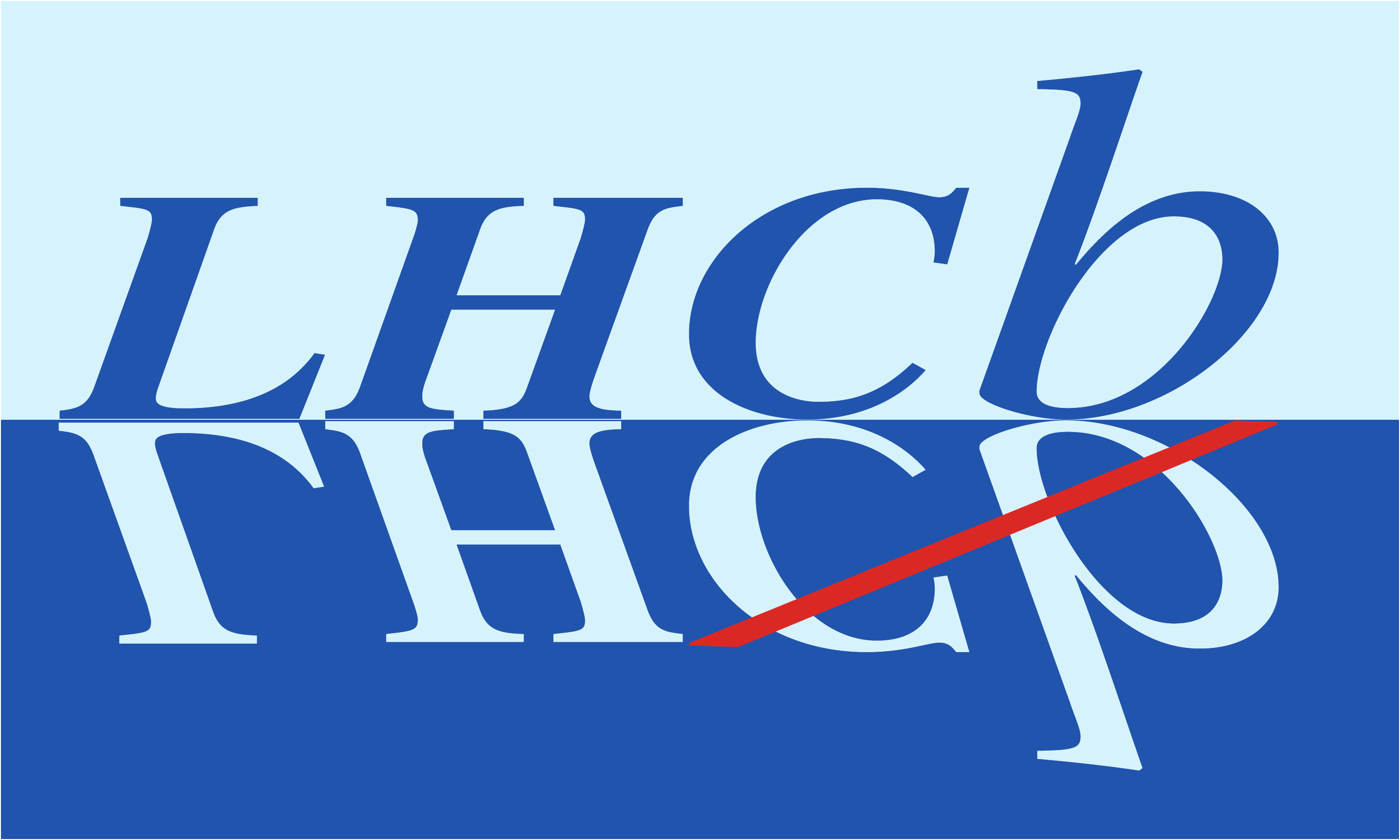}} & &}%
{\vspace*{-1.2cm}\mbox{\!\!\!\includegraphics[width=.12\textwidth]{figs/lhcb-logo.eps}} & &}%
\\
 & & CERN-EP-2022-101 \\  
 & & LHCb-PAPER-2021-050 \\  
 & & \today \\ 
 & & \\
\end{tabular*}

\vspace*{3.0cm}

{\normalfont\bfseries\boldmath\huge
\begin{center}
  \papertitle 
\end{center}
}

\vspace*{2.0cm}

\begin{center}
\paperauthors\footnote{Authors are listed at the end of this paper.}
\end{center}

\vspace{\fill}

\begin{abstract}
\noindent Measurements of \CP asymmetry in charmless $B\to PV$ decays are presented, where $P$ and $V$ denote a pseudoscalar and a vector meson, respectively. Five different $B\to PV$ decays from four final states, \pipipi, \kpipi, \kkk and \kkpi are analyzed. The measurements are based on a method that does not require full amplitude analyses, and are performed using proton-proton collision data at a center-of-mass energy of 13\,TeV collected by LHCb between 2015 and 2018, corresponding to an integrated luminosity of 5.9$\invfb$. In the $\pip\pim$ \textit{P}-wave, in the region dominated by the $B^\pm \to \rhoI K^\pm$ decay, a \CP asymmetry of \mbox{\acp = +0.150 $\pm$ 0.019 $\pm$ 0.011} is measured, where the first uncertainty is statistical and the second is systematic. This is the first observation of \CP violation in this process. For the other four decay channels, in regions dominated by the $B^\pm \to \rhoI \pi^\pm$, $B^\pm \to \kstar \pi^\pm$, $B^\pm \to \kstar K^\pm$ and $B^\pm \to \phiI K^\pm$ decays, \CP asymmetries in the \textit{P}-wave compatible with zero are measured. 

\end{abstract}

\vspace*{2.0cm}

\begin{center}
  Published in Phys. Rev. D108 (2023) 012013
\end{center}

\vspace{\fill}

{\footnotesize 
\centerline{\copyright~\papercopyright. \href{\paperlicenceurl}{\paperlicence}.}}
\vspace*{2mm}

\end{titlepage}


\newpage
\setcounter{page}{2}
\mbox{~}
%
%
%
%

\renewcommand{\thefootnote}{\arabic{footnote}}
\setcounter{footnote}{0}

\cleardoublepage


\pagestyle{plain} 
\setcounter{page}{1}
\pagenumbering{arabic}



\section{Introduction}
\label{sec:introduction}

In recent years, the large datasets produced at the LHC\@ have allowed precise measurements of direct \CP violation in $B$ meson decays~\cite{BediagaGobel}. However, there are still a number of decay channels without precise \CP$\,$-asymmetry measurements. Large samples of specific decays are required to improve our knowledge of \CP asymmetries in charmless decays of $B$ mesons, including those with neutral mesons in the final state. The start of Belle II~\cite{BelleII} operations, the coming data-taking with an upgraded LHCb detector~\cite{LHCb-PII-Physics}, and the analyses of the data already collected by the LHCb detector will allow the necessary measurements in the near future.

Theoretical developments using different approaches have resulted in many predictions for \CP asymmetries. Many of these studies are focused on charmless two-body and quasi-two-body \B-meson decays, in particular those to two pseudoscalar mesons ($B \to PP$) and to a pseudoscalar and a vector meson ($B \to PV$)~\cite{Beneke2000,Beneke2003,Smith2003,Soni2005,Zupan2006,Suzuki2007,Cheng2015,Beneke2015,Pat2016,Zhou2017,Cheng2020}. These studies are  directly linked to the long-standing controversy about the role of the short- and long-distance contributions to the generation of the strong-phase differences needed for direct \CP violation to occur~\cite{BediagaGobel}.
 
In this paper quasi-two-body $B\to PV$ decays, which result in three-body final states due to $V$ decays, are studied. Given the large phase space of these $B$-meson decays, different types of resonant contributions are allowed. Therefore, in three-body final states the vector resonances interfere with other resonant components. The interference has been used to estimate the strong phases, as well as the contribution from penguin amplitudes~\cite{Bigi1,Bigi2015,Grossman2021}. Furthermore, the three-body environment can affect the amount of the \CP violation associated with the $B \to PV$ decay amplitude~\cite{Pat2016}.

The resonant structure of three-body decays can be studied with model-dependent amplitude analyses, a complex task given the large number of possible intermediate states. Recently, the LHCb Collaboration presented three \CP $\,$-asymmetry measurements of  $B \to PV$ decays based on a full Dalitz plot analysis of data collected in 2011 and 2012~\cite{LHCb-PAPER-2017-033,LHCb-PAPER-2018-051,LHCb-PAPER-2019-017,LHCb-PAPER-2019-018}. From the amplitude analysis of the $B^0 \to K^0_S \pi^+ \pi^-$ decay, the \CP asymmetry in the \mbox{$B^0 \to K^*(892)^+ \pi^{-}$} decay was measured to be \acp = $-0.308 \pm 0.062$~\cite{LHCb-PAPER-2017-033}. 
The \acp measurement of the decay $B^\pm \to \kstar K^\pm$ from $B^\pm\to K^{\pm} \pi^{\mp}K^\pm$ decay was $0.123 \pm 0.087 \pm 0.045$ ~\cite{LHCb-PAPER-2018-051}. Finally, from the analysis of the $B^- \to \pi^-\pi^+\pi^-$ decay, the \CP asymmetry of $B^- \to \pi^- \rhoI$ decay was found to be consistent with zero, \mbox{\acp = $0.007 \pm 0.019$}~\cite{LHCb-PAPER-2019-017,LHCb-PAPER-2019-018}. In the same analysis, \CP asymmetries were observed in the $B^- \to \pi^- \sigma$ and $B^- \to \pi^- f_2(1270)$ decays, \mbox{\acp  $=0.160 \pm 0.028$} and \acp $=0.468 \pm 0.077$, respectively.

In this paper, measurements of \CP asymmetries in charmless $B \to PV$ decays are presented. The measurements are based on data collected by the LHCb detector between 2015 and 2018, corresponding to 5.9\invfb of proton-proton ($pp$) collisions at a center-of-mass energy of 13\,TeV. A new method~\cite{Pat2016} that does not rely on a full amplitude analysis is used. The method is based on three key features of three-body $B$ decays: the large phase space; the dominance of scalar and vector resonances with masses below or around 1\gevcc, confirmed by amplitude analyses performed by Belle~\cite{PhysRevLett.96.251803,Belle2010}, BaBar~\cite{Babar2005,Babar2011,Babar2012,Babar2015} and LHCb Collaborations~\cite{LHCb-PAPER-2019-017,LHCb-PAPER-2019-018,LHCb-PAPER-2018-051}; and the clear signatures of the resonant amplitudes in the Dalitz plot. The method used in this analysis is suited for measuring the \CP asymmetry between the yields of the $B^+\to P^+V$ and $B^-\to P^-V$ decays.  

In the decay $B^\pm \to R(\to h_1^-h_2^+)h_3^\pm$, where $R$ is a resonance, the notation $s_{\parallel}$ is used for the two-body invariant mass squared $m^2(h_1^-h_2^+)$ and \sperp for $m^2(h_1^-h_3^+)$. The resonance line shape (typically a Breit-Wigner distribution) is observed in the projection of the Dalitz plot onto the $s_{\parallel}$ axis. When a narrow interval in $s_{\parallel}$ around the resonance mass  is selected, the projection of the data onto $s_{\perp}$ reflects the angular distribution of the decay products. In vector resonances, a parabolic shape is observed, since the decay width is proportional to cosine squared of the helicity angle, $\cos^2\theta$, where $\theta$ is defined as the angle between $h^-_1$ and $h^+_3$ computed in the ($h^-_1,h^+_2$) rest frame. If the ($h^-_1,h^+_2$) pair forms a scalar resonance, the distribution in $s_{\perp}$ is uniform, since the decay of scalar resonances is isotropic in $\cos\theta$. The interference term between a vector and a scalar resonance is linear in $\cos\theta$. 

The \CP asymmetry is measured for the following decays: the $B^\pm \to \rhoI K^\pm $ region and $B^\pm \to \KorKbar^*(892)^0 \pi^\pm$ from the \kpipi final state; $B^\pm \to \phiI K^\pm $ from \kkk decays; the $B^\pm \to \rhoI \pi^\pm$ region from the \pipipi final state; and $B^\pm \to \KorKbar^*(892)^0 K^\pm$ from \kkpi decays. It is important to emphasize that the method does not isolate the \rhoI contribution from the influence of the \omegaI resonance. However, previous analysis show that the fit fractions of amplitudes involving the \rhoI resonances are roughly two orders of magnitude higher than those of the \omegaI in the $B^\pm \to \rhoI \pi^\pm$ and $B^\pm \to \rhoI K^\pm$ decays~\cite{LHCb-PAPER-2019-018,BaBar_bkpipi}. In addition, their widths are about one order of magnitude different. Therefore, hereafter the \textit{P}-wave decays in the regions dominated by the \rhoI vector resonance will be denoted as $B^\pm \to \rhoI \pi^\pm $ and $B^\pm \to \rhoI K^\pm $.

The method introduced in~\cite{Pat2016} is described in Sec.~\ref{sec:strategy}. A detailed description of the selection, efficiency and background for the four charmless three-body channels is given in a companion paper~\cite{LHCb-PAPER-2021-049}.


\section{LHCb detector and dataset}
\label{sec:detector}

The LHCb detector \cite{LHCb_detector2008,LHCb-DP-2014-002} is a single-arm forward spectrometer covering the pseudorapidity range $2 < \eta < 5$, designed for the study of particles containing $b$ or $c$ quarks. The detector includes a high-precision tracking system consisting of a silicon-strip vertex detector surrounding the $pp$ interaction region, a large-area silicon-strip detector located upstream of a dipole magnet with a bending power of about 4 Tm, and three stations of silicon-strip detectors and straw drift tubes placed downstream of the magnet. The tracking system provides a measurement of the momentum, $p$, of charged particles with a relative uncertainty that varies from 0.5$\%$ at low momentum to 1.0$\%$ at 200\gevc. The minimum distance of a track to a primary $pp$ collision vertex, the impact parameter (IP), is measured with a resolution of (15 + 29/$p_\text{T}$)\,$\upmu$m, where $p_\text{T}$ is the component of the momentum transverse to the beam, in\,\gevc. Different types of charged hadrons are distinguished using information from two ring-imaging Cherenkov detectors. Photons, electrons and hadrons are identified by a calorimeter system consisting of scintillating- pad and preshower detectors, an electromagnetic and a hadronic calorimeter. Muons are identified by a system composed of alternating layers of iron and multiwire proportional chambers.

The online event selection is performed by a trigger, consisting of a hardware stage, based on information from the calorimeter system, followed by a software stage, which applies a full event reconstruction. At the hardware trigger stage, the $B^+\to h_1^-h_2^+h_3^+$ candidates are required to include a hadron with transverse energy deposited in the calorimeters typically larger than 3.5\gev. The software trigger requires a two-, three- or four-track vertex with a significant displacement from all primary vertices. At least one charged particle must have a large transverse momentum and be inconsistent with originating from any primary vertex. A multivariate algorithm is used for the identification of displaced vertices consistent with the decay of a {\it b}-hadron.

Simulations are used to model the effects of the detector acceptance and the selection requirements, to validate the fit models and to evaluate efficiencies. In the simulation, $pp$ collisions are generated using \pythia 8 \cite{Sj_strand_2015} with a specific LHCb configuration \cite{LHCb-PROC-2010-056}. Decays of unstable particles are described by \evtgen \cite{Lange:2001uf}, in which final-state radiation is generated using \photos \cite{davidson2015photos}. The interaction of the generated particles with the detector, and its response, are implemented using the \geant toolkit \cite{Agostinelli:2002hh,Allison:2006ve} as described in Ref. \cite{LHCb-PROC-2011-006}.


\section{Selection of signal candidates}
\label{sec:selection}

The selection of signal candidates follows closely the procedure used in the model-independent analysis of the same data sample \cite{LHCb-PAPER-2014-044}. Signal $B^+$ candidates are formed from three tracks that are consistent with originating from the same secondary vertex. Each reconstructed $B^+$ candidate is associated with the primary vertex that is most consistent with its flight direction. A requirement is also imposed on the angle between the $B^+$ momentum and the vector between the primary and secondary vertices.

A multivariate analysis is performed to further reduce the combinatorial background. A
boosted decision-tree classifier \cite{Breiman} is trained using simulated signal and data in the high-mass sideband region ($m_B > 5.4$\gevcc) for the background. The variables used in this classifier are the
quantities based on the quality of the reconstructed tracks and decay vertices, the kinematic properties of the
$B^+$ candidate and its decay products, and the $B^+$ candidate displacement from the primary vertex. 
The requirement on the response of this classifier is chosen to optimize the statistical significance of the signal, $\varepsilon_{\textrm{sim}}/\sqrt{(S+B)_{\textrm{data}}}$, where $\varepsilon_{\textrm{sim}}$ is the signal efficiency determined in simulation and $(S+B)_{\textrm{data}}$ is obtained by counting the events selected from data within $\pm$40\mevcc  of the known $B^+$ mass~\cite{PDG2020}.

Particle identification (PID) is used to reduce the cross-feed from other $B$ decays in which hadrons are incorrectly identified. The main sources of this cross-feed are $K \to \pi$ and $\pi \to K$ misidentification.
These backgrounds arising from $K \to \pi$ and $\pi \to K$ misidentification are suppressed by stringent PID requirements for each final-state particle.
Tracks that are outside of the fiducial region of the PID system are removed. Furthermore, tracks associated with hits in the muon system are  removed to eliminate cross-feed from semileptonic decays.  

Candidates within the invariant mass interval 5247$-$5315~\mevcc, which includes approximately 95\% of the considered $B^{\pm}$ decays, are retained for further analysis. The number of $B^{\pm}$ candidates for each channel used in this analysis, as well as the signal purity, are shown in Table~\ref{tab:intSigReg}. 

\begin{table}[h!]
\centering
\caption{\small Number of $B^+$ and $B^-$ candidates in the signal region of 5247 to 5315\mevcc and the corresponding purities. }
\begin{tabular}{l|c|c|c|c}
    & \kpipi &  \kkk  & \pipipi & \kkpi \\
\hline
$B^-$  & 243\,960 & 159\,673 & 51\,977 & 17\,161 \\ 
$B^+$  & 240\,884 & 176\,345 & 44\,389 & 21\,178 \\
Purity & 0.91 & 0.96 & 0.88 & 0.76 \\
\end{tabular}
\label{tab:intSigReg}
\end{table} 

From these candidates, vector resonances are selected by applying restrictions on the $s_{\parallel}$ variable around the known mass of each involved resonance, i.e., \kstar, \rhoI and \phiI. In the $s_{\parallel}$ axis, the data are analyzed in invariant mass intervals of 50, 150 and 5 \mevcc, respectively, centered at the known values of the resonance masses. Since all decay modes have resonances in both $s_\parallel$ and \sperp, only data with $\sperp > 5$ \gevgevcccc are considered. This requirement ensures that only the interference between scalar and vector resonances in $s_\parallel$ is relevant. The definition of the interval in \sperp varies according to the position of the resonance in the phase space. In order to avoid charmonium resonances in the $\pi^+ \pi^-$ spectrum of the \kpipi decay, an additional veto is applied around the known $\chi_{c0}$ and $J/\psi$ invariant masses.


\section{\boldmath{$B \to PV$} fit function}
\label{sec:strategy}
 
Generally, the decay amplitudes for $B^+$ and $B^-$ are represented as a coherent sum of intermediate amplitudes, with the magnitude and the phase for each amplitude as free parameters. At low two-body invariant masses the data are dominated by scalar and vector
resonances. In the case of one vector resonance interfering with a scalar component, the decay amplitudes can be represented by~\cite{Pat2016}
 
\begin{equation}
	\begin{aligned}
		\mathcal{M}_{\pm} = a_{\pm}^V e^{i \delta_{\pm}^V}F_V^{\rm BW} \cos \theta(s_\bot,s_\|) + a_{\pm}^{\rm S} e^{i \delta_{\pm}^{S}}F_{S}^{\rm BW}, \\
	\end{aligned}
	\label{eq:M+M-}
\end{equation} 
where $a^V_{\pm}$ and $a^S_{\pm}$ are the magnitudes of the vector and  scalar resonances, respectively, assumed to be independent of \sperp. $\delta_{\pm}^{V}$ and $\delta_{\pm}^{S}$ are the phases of the vector and scalar amplitudes, and $\theta(s_\bot,s_\|)$ is the helicity angle. The resonance $R$ may be described by a Breit-Wigner (BW) function, $F^{\rm BW}_{R}$, without any loss of generality, 

\begin{eqnarray}
F^{\rm  BW}_{R} (s_{\parallel}) = \frac{1}{m^2_{R} - s_{\parallel} - i m_{R}\Gamma_{R}(s_{\parallel})} ~,
\label{eq:BW}
\end{eqnarray}
where $\Gamma_{R}(s_{\parallel})$ is the energy-dependent relativistic width and $m_R$ is the resonance mass. The helicity angle is a function of the two Dalitz variables, $\cos\theta(s_{\parallel}, s_\perp)$~\cite{Byckling:1971vca}. However, for the low mass and sufficiently narrow resonances, a parabolic dependence of $\cos\theta$ only on $s_\perp$ is a good approximation. 

The matrix element squared is
\begin{equation}
  \begin{aligned}
    \left|\mathcal{M}_{\pm}\right|^{2} &= (a_{\pm}^{V})^{2} (\cos \theta)^{2} |F_{V}^{\rm BW}|^{2} + (a_{\pm}^{S})^{2} |F_{S}^{\rm BW}|^{2} + 2 a_{\pm}^{V} a_{\pm}^{S} \cos \theta |F_{V}^{\rm BW}|^{2} |F_{S}^{\rm BW}|^{2} \\ 
    & \times 
    \{ \cos (\delta_{\pm}^{V} - \delta_{\pm}^{S}) [(m_{V}^{2} - s_\|) (m_{S}^{2} - s_\|) + (m_{V}\Gamma_{V}) (m_{S}\Gamma_{S})]\ \\
    & +
    \sin (\delta_{\pm}^{V} - \delta_{\pm}^{S}) [(m_{S}\Gamma_{S}) (m_{V}^{2} - s_\|) - (m_{V}\Gamma_{V}) (m_{S}^{2} - s_\|)] \},
  \end{aligned}
\label{eq:ampiv}
\end{equation}
where $m_{V(S)}$ is the vector (scalar) mass and the dependency of $\cos\theta$ on $s_{\parallel}$ and \sperp is omitted for simplicity.

Assuming that $a^V_{\pm}$, $a^S_{\pm}$ and the phases $\delta_{\pm}^V$ and $\delta_{\pm}^S$ do not depend on \sperp, Eq.~\eqref{eq:ampiv} can be simplified as a quadratic polynomial in $\cos\theta(m_V^2,\sperp)$ and written as  
\begin{equation}
	|\mathcal{M_\pm}|^2 = f(\cos\theta(m_V^2,\sperp)) = \pzero^\pm + \pone^\pm \cos\theta(m_V^2,\sperp) + \ptwo^\pm \cos^2\theta(m_V^2,\sperp),
	\label{eq:quadfit}
\end{equation} 
where $p^{\pm}_{0,1,2}$ are polynomial coefficients. The coefficient $p^{\pm}_{0}$   is related to \CP violation in the the scalar component, whereas $p^{\pm}_{1}$ is related to \CP violation in the interference between the vector and scalar amplitudes. Since the scalar resonances are usually broad, an amplitude analysis is required to quantify the \CP violation in these cases. 
Given that the decay rates are proportional to $|\mathcal{M_\pm}|^2$, the \CP asymmetry $A^V_{\CP}$ in the \BPV decay is given as function of $\ptwo^\pm$,

\begin{equation}
A^V_{\CP} =  \frac{|\mathcal{M_-}|^2  - |\mathcal{M_+}|^2 }{|\mathcal{M_-}|^2  + |\mathcal{M_+}|^2 } = \frac{\ptwo^-  - \ptwo^+ }{\ptwo^- + \ptwo^+}.
\label{eq:Acp_fit}
\end{equation}
Given the approximation $\cos\theta(s_{\parallel},s_{\perp}) \simeq \cos\theta(m_V^2,s_{\perp})$, $\cos\theta$ becomes a linear function of $s_{\perp}$~\cite{Byckling:1971vca}. With this approximation, the \CP asymmetry can be obtained from the distribution of \sperp, whereas the asymmetry obtained from the $\cos\theta$ distribution is used to evaluate the systematic uncertainty.

Finally, the function in Eq.~\eqref{eq:quadfit} is used to fit the histograms of data projected onto the $s_{\perp}$ axes in order to determine the fit parameters $p^{\pm}_{0,1,2}$, and then calculate the resulting \CP asymmetry using Eq.~\eqref{eq:Acp_fit}.
The asymmetry depends only on the first term of Eq.~\eqref{eq:ampiv}, related to $\cos^2 \theta$. The other terms in this equation are constant or linearly dependent on $\cos\theta$. 


\section{Results}
\label{sec:Results}

\begin{figure}[tb]
\centering
\begin{overpic}[width=0.49\linewidth]{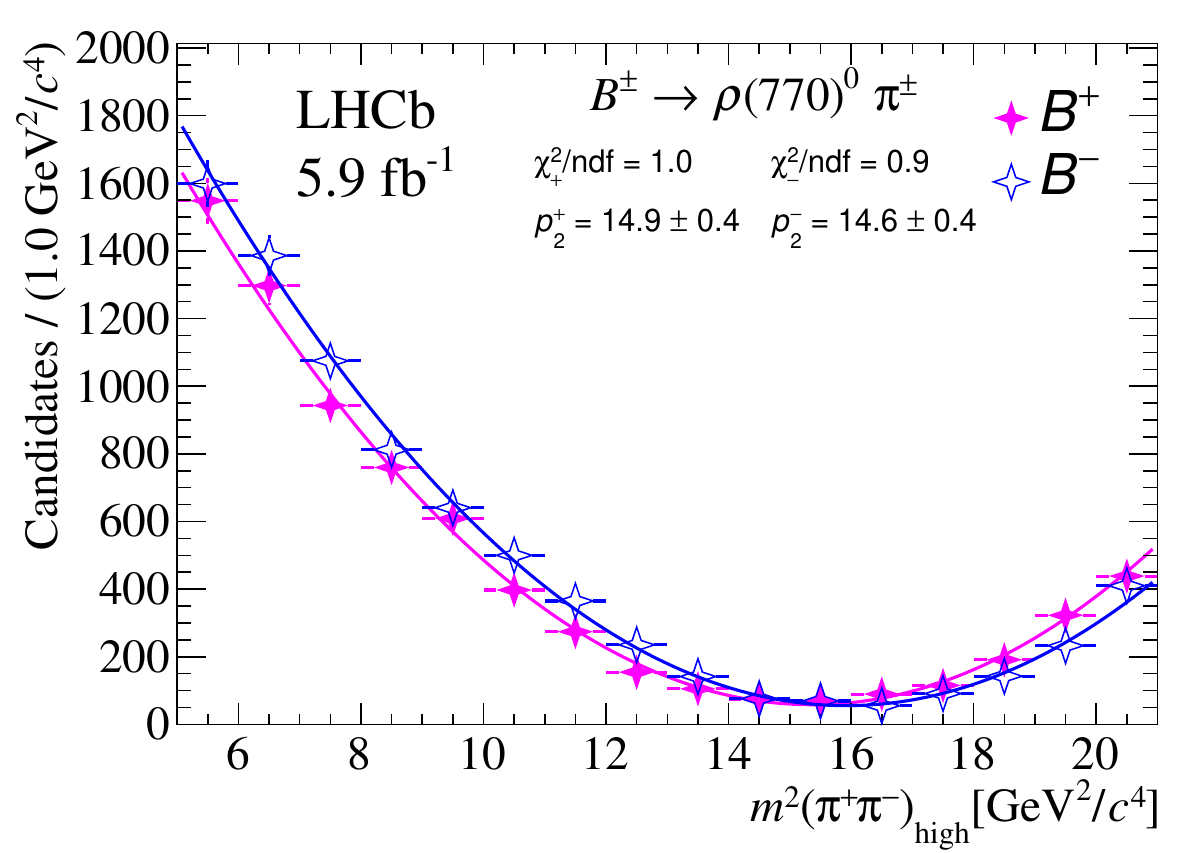}
 \put(18,63){\scriptsize{(a)}}
\end{overpic}
\begin{overpic}[width=0.49\linewidth]{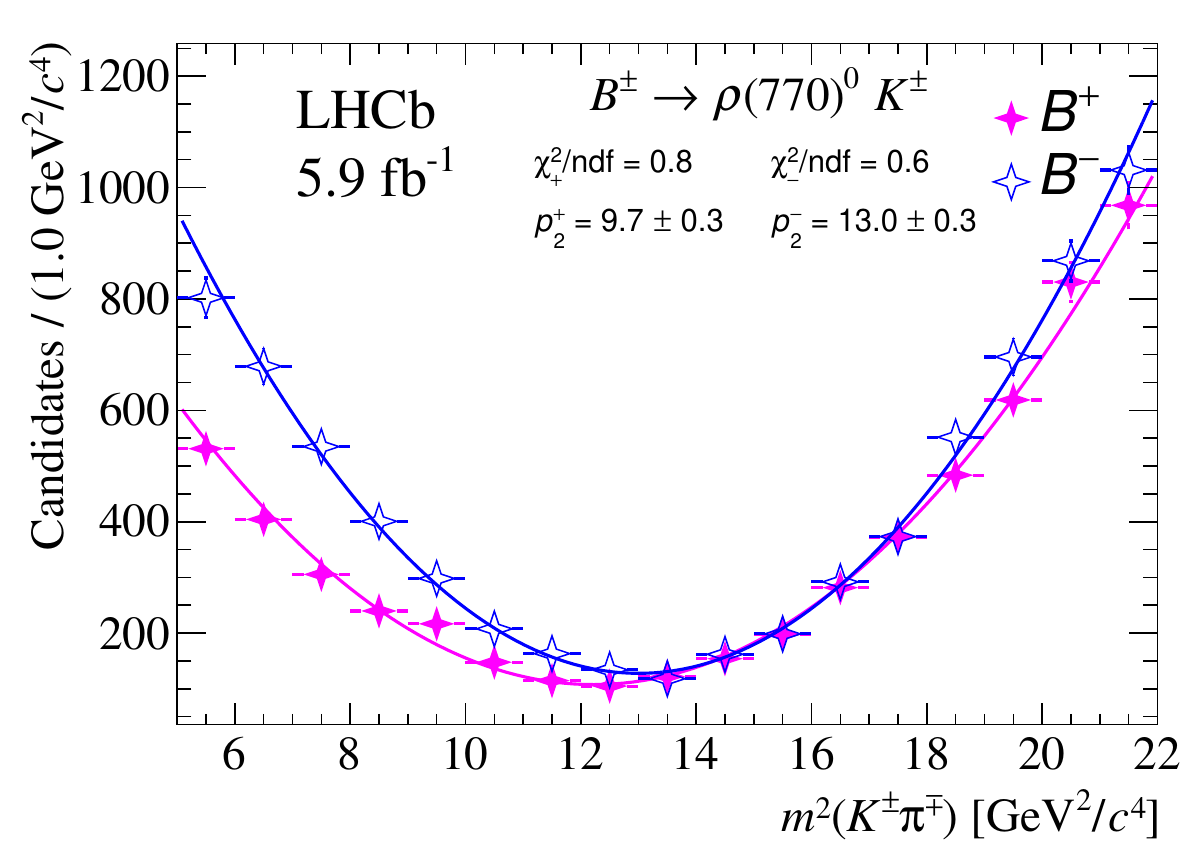}
 \put(18,63){\scriptsize{(b)}}
\end{overpic}

\vspace{0.2cm}

\begin{overpic}[width=0.49\linewidth]{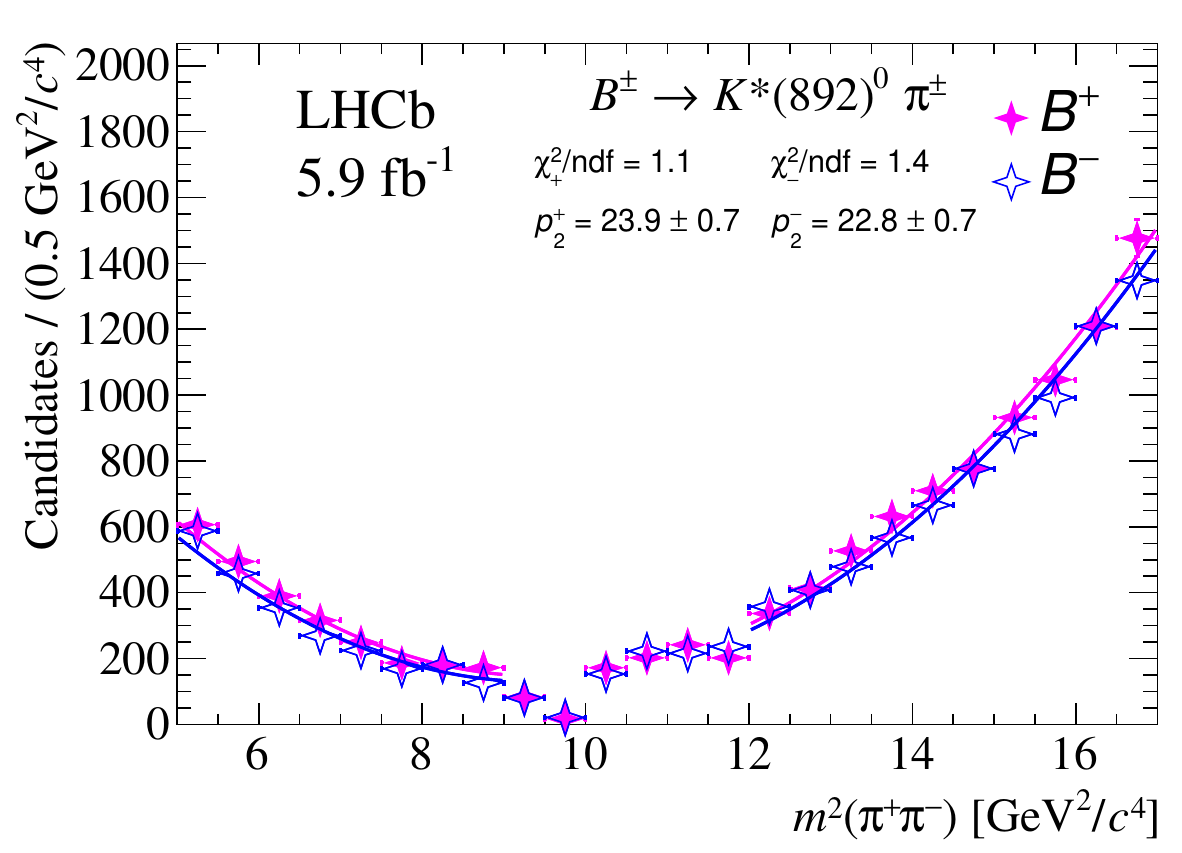}
 \put(18,63){\scriptsize{(c)}}
\end{overpic}
\begin{overpic}[width=0.49\linewidth]{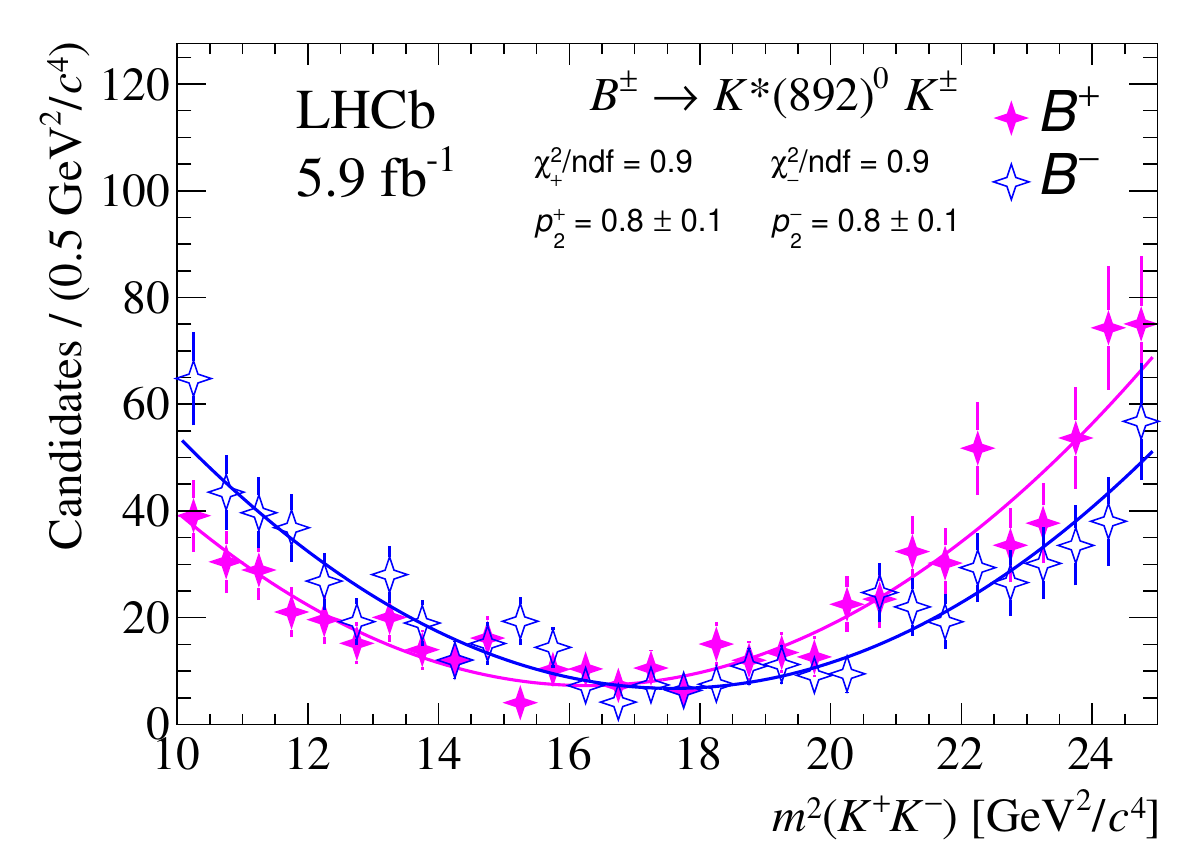}
  \put(18,63){\scriptsize{(d)}}
\end{overpic}

\vspace{0.2cm}

\begin{overpic}[width=0.49\linewidth]{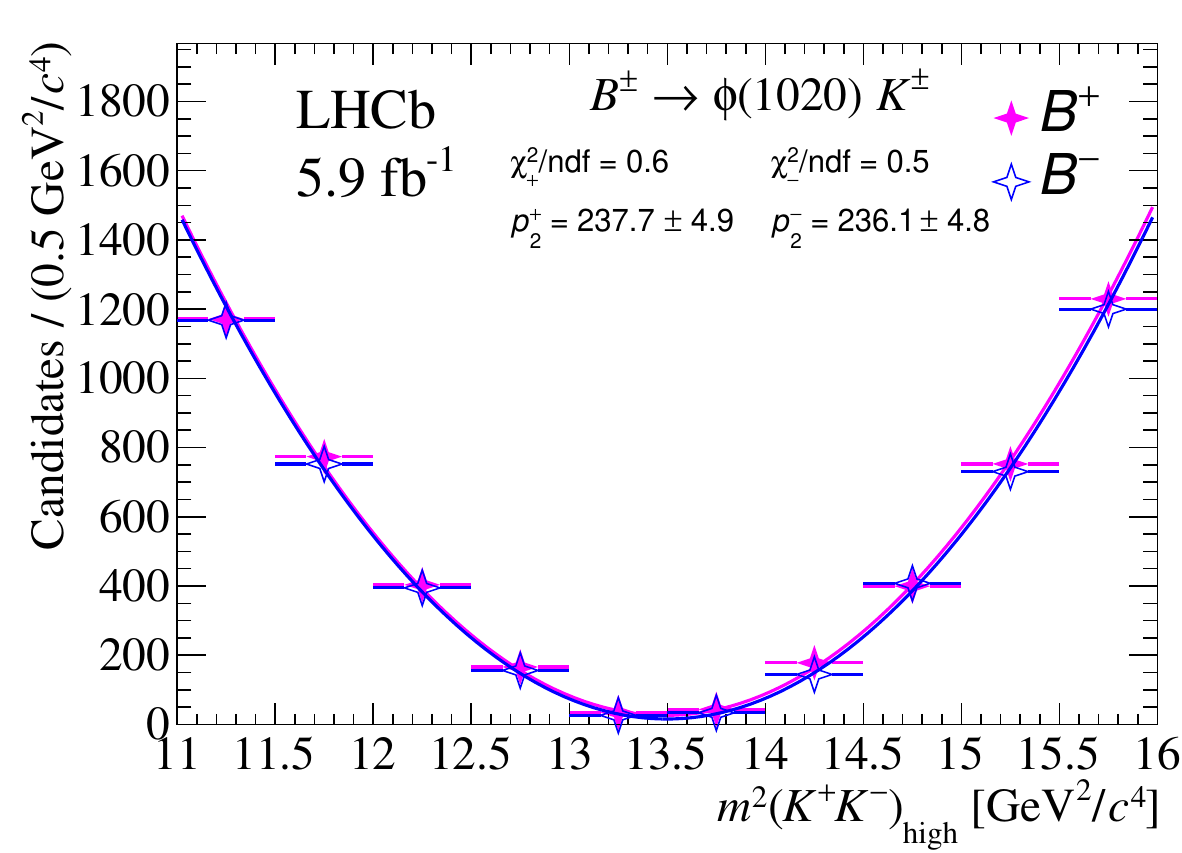}
  \put(18,63){\scriptsize{(e)}}
\end{overpic}

\caption{Distribution of \sperp for $B^+$ and $B^-$ candidates and the corresponding quadratic fits for (a) \rhoI in \pipipi, (b) \rhoI in \kpipi, (c) \kstar in \kpipi, (d) \kstar in \kkpi and (e) \phiI in \kkk. In the symmetric channels, the phase space distribution and its projections are presented with the two axes being the squares of the low-mass $m_{\textrm{low}}$ and high-mass $m_{\textrm{high}}$ combinations of the opposite-sign particle pairs, for visualization purposes.}
\label{fig:fitm2}
\end{figure}

The efficiency-corrected yields of \Bp and \Bm as a function of \sperp are displayed in Fig.~\ref{fig:fitm2}, with the results of the quadratic fits [Eq.~\eqref{eq:quadfit}] superimposed. The fit parameters, as well as the corresponding goodness-of-fit parameter \chisqndf, are also listed. All asymmetries are computed from the term $p_{2}^{\pm}$ and corrected for the \Bp-meson production asymmetry~\cite{LHCb-PAPER-2021-049}.

The vector resonances studied in this paper occupy a small part of the charmless three-body $B$ phase-space decay. The combinatorial background behavior in this region is a smooth function of the Dalitz variables, so the parameters $p_0^\pm$ and $p_1^\pm$ absorb it. Another background component is related to the prompt production of these resonances plus a random track. It has an angular distribution similar to the scalar resonances, so it is absorbed in the $p_0^\pm$ parameter.

\subsection{$\boldsymbol{\pipipi}$ decay}

For the $\Bp \to \rhoI \pip$ region, the \CP asymmetry related to the vector resonance is measured to be 

\begin{equation}
	\acp(\rhoI \pi^\pm) = -0.004 \pm 0.017, \nonumber
\end{equation} which is compatible with \CP symmetry. 

The effect of a \CP asymmetry compatible with zero can also be seen in Fig.~\ref{fig:fitm2} (a), which shows the vector parabolas  of \Bp and \Bm very close to each other. It is important to note that, given the mass window selected, this measurement also includes the \omegaI contribution.

\subsection{\boldmath{\kpipi} decay}

This decay has two amplitudes involving low-mass vector resonances: $B^\pm \to \kstar \pi^\pm$ and the region dominated by $B^\pm \to \rhoI K^\pm$ decays.
Unlike the result for the \mbox{$\Bp \to \rhoI \pip$} region, the large \CP asymmetry obtained here for the $\Bp \to \rhoI \Kp$ region can be clearly seen as a difference between the \Bp and \Bm parabolas in Fig.~\ref{fig:fitm2} (b). In this \textit{P}-wave region dominated by the \rhoI resonance, the \CP asymmetry is measured to be

\begin{equation}
	\acp(\rhoI K^\pm) = 0.150 \pm 0.019, \nonumber
\end{equation} 
with a statistical-only significance of 7.9 standard deviations ($\sigma$). This measurement can be compared with the values obtained by previous experiments for this channel, as listed in Table \ref{tab:systematics_bpv}. The results are compatible within the uncertainties.

For $\Bp \to \kstar \pip$ decays, the region with $\pi^+ \pi^-$ mass in the range [9, 12]~\gevgevcccc is removed from the fit due to the presence of the \jpsi and \chiczero resonances, as can be seen in Fig.~\ref{fig:fitm2} (c). The \CP asymmetry related to the vector resonance is measured to be   

\begin{equation*}
	\acp(\kstar \pi^\pm) = -0.015 \pm 0.021, \nonumber
\end{equation*} 
which is compatible with \CP symmetry.

The exclusion of narrow regions around the charmonium resonances does not affect the sensitivity of the method or the fit quality, since these regions are close to the minimum of the parabolas. Again, the similarity between the \Bp and \Bm parabolas in Fig.~\ref{fig:fitm2} (c) is compatible with the numerical value obtained.

\subsection{\boldmath{\kkpi decay}}

For the $\Bp \to \kstar \Kp$ resonance, the data are analyzed in the range [10, 25]~\gevgevcccc, due to the smaller phase space of the \kkpi final state.

The \CP asymmetry related to the vector resonance, shown in Fig.~\ref{fig:fitm2} (d), is measured to be  

\begin{equation*}
	\acp(\kstar K^\pm) = 0.007 \pm 0.054,
	\label{eq:kstar-kkpi-yield}
\end{equation*} 
which is compatible with \CP symmetry.

\subsection{\boldmath{\kkk decay}}

For $\Bp \to \phiI \Kp$ decays, the \CP asymmetry related to the vector resonance is measured to be  

\begin{equation*}
	\acp(\phiI K^\pm) = 0.004 \pm 0.014,
	\label{eq:phi-kkk-yield}
\end{equation*} 
consistent with \CP symmetry. This measurement is in agreement with the value obtained by the \babar experiment~\cite{Babar2012}, $\ACPphi = +0.128 \pm 0.044 \pm 0.013$, at the 2.8 $\sigma$ level.
The similarity of the \Bp and \Bm distributions in Fig.~\ref{fig:fitm2} (e) is consistent with the small value of \ACPphi obtained.


\section{Systematic uncertainties}
\label{sec:systematics}

The three leading sources of systematic uncertainties are discussed below, the dominant one being the variation of the range in \sperp where the fits are performed.

\textbf{Variation of fit regions:} The range in \sperp where the data are fitted varies according to the phase space and the presence of other resonances.
The default values of projections, in units of \gevgevcccc, are 5$-$21 for the \rhoI resonance in \pipipi decays, 5$-$22 for \rhoI in \kpipi decays, 5$-$17 for \kstar in \kpipi decay, 10$-$25 for \kstar in \kkpi decay and 11$-$16 for \phiI in \kkk decay. 
The intervals are varied by displacing simultaneously both low and high limits by up to 0.5 \gevgevcccc for \phiI in \kkk decay, and by up to 1 \gevgevcccc for all other decays.

\textbf{Variations of resonance mass window:} The choice of interval in $s_{\parallel}$ around the resonance mass defines the region where the data are fitted.
The intervals in $s_{\parallel}$ are varied around the default values, described in the Sec.~\ref{sec:selection}, considering the ranges 140$-$160, 45$-$55 and 4.5$-$5.5~\mevcc for \rhoI, \kstar and \phiI, respectively. The differences in the results with respect to the default fit are taken as systematic uncertainties in the corresponding \CP-asymmetry measurements. The variation is done in small increments, giving 1000 results for each channel. The systematic uncertainties are taken from the root mean square of the resulting asymmetry distributions. 

\textbf{Change of the projected variable:} In this case, the fit is performed defining the parabola in terms of the helicity angle $\cos\theta$, instead of \sperp. The procedure to obtain the \CP asymmetry is the same and the difference with respect to the default fit is taken as systematic uncertainty.

The need for higher-order terms in the fit function is also investigated. These terms would account for a possible influence of $f_2(1270)$ in the \pipipi final state. Using simulation~\cite{Back:2017zqt} and the known value of $\mathcal{B}(f_2(1270)\to\pi^+\pi^-$), the contribution of the tensor resonance is found to be negligible. Finally, a systematic uncertainty related to the efficiency correction was evaluated and also found to be negligible.

\begin{table}[tb]
\footnotesize
\centering
\caption{Summary of \CP-asymmetry measurements for the vector resonance channels and their associated final-state $B^\pm \to R(\to h_1^-h_2^+)h_3^\pm$ decays. For comparison purposes, the previous measurements from other experiments are also included. }
\begin{tabular}{lcl}

\toprule
 
Decay channel  & This work & Previous measurements \\ 
\midrule
$B^\pm \to$(\rhoI $\to \pi^+ \pi^-$)$\pi^\pm$        & $-$0.004 $\pm$ 0.017 $\pm$ 0.009 & $+$0.007 $\pm$ 0.011 $\pm$ 0.016 (\lhcb \cite{LHCb-PAPER-2019-017,LHCb-PAPER-2019-018})  \\ \midrule
$B^\pm \to$(\rhoI $\to \pi^+ \pi^-$)$K^\pm$        & $+$0.150 $\pm$ 0.019 $\pm$ 0.011 & $+$0.44  $\pm$ 0.10  $\pm$ 0.04  (\babar~\cite{BaBar_bkpipi}) \\
		                                &                                  & $+$0.30  $\pm$ 0.11  $\pm$ 0.02  (\belle~\cite{PhysRevLett.96.251803}) \\ \midrule
$B^\pm \to$(\kstar $\to K^{\pm} \pi^{\mp}$)$\pi^\pm$ & $-$0.015 $\pm$ 0.021 $\pm$ 0.012 & $+$0.032 $\pm$ 0.052 $\pm$ 0.011 (\babar~\cite{BaBar_bkpipi}) \\ 
		                                &                                  & $-$0.149 $\pm$ 0.064 $\pm$ 0.020 (\belle~\cite{PhysRevLett.96.251803}) \\ \midrule
$B^\pm \to$(\kstar $\to K^{\pm} \pi^{\mp}$)$K^\pm$ & $+$0.007 $\pm$ 0.054 $\pm$ 0.032 & $+$0.123 $\pm$ 0.087 $\pm$ 0.045 (\lhcb~\cite{LHCb-PAPER-2018-051})  \\ \midrule
$B^\pm \to$(\phiI  $\to K^+ K^-$)$K^\pm$           & $+$0.004 $\pm$ 0.014 $\pm$ 0.007 & $+$0.128 $\pm$ 0.044 $\pm$ 0.013 (\babar~\cite{Babar2012}) \\ 
\bottomrule

\end{tabular}

\label{tab:systematics_bpv}
\end{table}

The total systematic uncertainties are obtained as the sum in quadrature of the three contributions. Table \ref{tab:systematics_bpv} summarizes the results obtained in this analysis.


\section{Summary and conclusion}
\label{sec:conclusion}

In summary, \CP asymmetries in charmless \BPV decays are determined using a new method, without the need for amplitude analyses. The data set analyzed corresponds to an integrated luminosity of 5.9\invfb of proton-proton collisions collected by the LHCb detector in 2015$-$2018 at a center-of-mass energy of 13 TeV. Five decay channels are studied, namely $B^\pm \to \phi(1020)K^\pm$, $B^\pm \to \kstar \pi^\pm$, $B^\pm \to \rhoI \pi^\pm$, $B^\pm \to \kstar K^\pm$ and $B^\pm \to \rho(770)^0K^\pm$. For the $B^\pm \to \rho(770)^0K^\pm$ region, the \CP asymmetry is measured to be $\acp = +0.150 \pm 0.019 \pm 0.011$, which differs from zero by 6.8$\sigma$, computed with the total uncertainty.

For the other channels, the measured \CP asymmetries are compatible with zero, as predicted using the \CPT constraint\cite{Pat2016}. The \CPT symmetry would suppress \CP violation in \BPV decays, which nevertheless could still occur through final-state interactions involving the third particle. A distinct feature of the $B^\pm \to \rho(770)^0K^\pm$ amplitude in the  \kpipi final state is that the contribution from the vector amplitude is much smaller than the scalar contribution, represented by the $B^\pm \to f_0(980)^0K^\pm$ decay, whereas the opposite is true for the other final states studied.

These measurements are significantly more precise than the previous results obtained by the Belle and BaBar Collaborations. Some tension is found between the results of this analysis and those from Belle and BaBar, whereas good agreement is found with LHCb results obtained with amplitude analyses of $\pipipi$  \cite{LHCb-PAPER-2019-017, LHCb-PAPER-2019-018} and \kkpi decays \cite{LHCb-PAPER-2018-051}.

The method used in this analysis is based on the approximation of a two-body interaction plus one spectator meson, and on the general assumption that the magnitudes and  phases of the amplitudes are constant across the whole phase space. These hypotheses, which are assumed by all models used in amplitude analyses,  are supported by the quality of the fits.

\section*{Acknowledgments}
%
%
\noindent We express our gratitude to our colleagues in the CERN
accelerator departments for the excellent performance of the LHC. We
thank the technical and administrative staff at the LHCb
institutes.
We acknowledge support from CERN and from the national agencies:
CAPES, CNPq, FAPERJ and FINEP (Brazil); 
MOST and NSFC (China); 
CNRS/IN2P3 (France); 
BMBF, DFG and MPG (Germany); 
INFN (Italy); 
NWO (Netherlands); 
MNiSW and NCN (Poland); 
MEN/IFA (Romania); 
MICINN (Spain); 
SNSF and SER (Switzerland); 
NASU (Ukraine); 
STFC (United Kingdom); 
DOE NP and NSF (USA).
We acknowledge the computing resources that are provided by CERN, IN2P3
(France), KIT and DESY (Germany), INFN (Italy), SURF (Netherlands),
PIC (Spain), GridPP (United Kingdom), 
CSCS (Switzerland), IFIN-HH (Romania), CBPF (Brazil),
Polish WLCG  (Poland) and NERSC (USA).
We are indebted to the communities behind the multiple open-source
software packages on which we depend.
Individual groups or members have received support from
ARC and ARDC (Australia);
Minciencias (Colombia);
AvH Foundation (Germany);
EPLANET, Marie Sk\l{}odowska-Curie Actions and ERC (European Union);
A*MIDEX, ANR, IPhU and Labex P2IO, and R\'{e}gion Auvergne-Rh\^{o}ne-Alpes (France);
Key Research Program of Frontier Sciences of CAS, CAS PIFI, CAS CCEPP, 
Fundamental Research Funds for the Central Universities, 
and Sci. \& Tech. Program of Guangzhou (China);
GVA, XuntaGal, GENCAT and Prog. Atracci\'on Talento, CM (Spain);
SRC (Sweden);
the Leverhulme Trust, the Royal Society
 and UKRI (United Kingdom).




\addcontentsline{toc}{section}{References}
\bibliographystyle{LHCb}
\bibliography{main,standard,LHCb-PAPER,LHCb-CONF,LHCb-DP,LHCb-TDR}
 
\newpage

\centerline
{\large\bf LHCb collaboration}
\begin
{flushleft}
\small
R.~Aaij$^{32}$\lhcborcid{0000-0003-0533-1952},
A.S.W.~Abdelmotteleb$^{50}$\lhcborcid{0000-0001-7905-0542},
C.~Abellan~Beteta$^{44}$,
F.~Abudin{\'e}n$^{50}$\lhcborcid{0000-0002-6737-3528},
T.~Ackernley$^{54}$\lhcborcid{0000-0002-5951-3498},
B.~Adeva$^{40}$\lhcborcid{0000-0001-9756-3712},
M.~Adinolfi$^{48}$\lhcborcid{0000-0002-1326-1264},
H.~Afsharnia$^{9}$,
C.~Agapopoulou$^{13}$\lhcborcid{0000-0002-2368-0147},
C.A.~Aidala$^{77}$\lhcborcid{0000-0001-9540-4988}
% ...

{\footnotesize \it

% ...
$^{9}$Universit{\'e} Clermont Auvergne, CNRS/IN2P3, LPC, Clermont-Ferrand, France\\
% ...
$^{13}$LPNHE, Sorbonne Universit{\'e}, Paris Diderot Sorbonne Paris Cit{\'e}, CNRS/IN2P3, Paris, France\\
% ...
$^{32}$Nikhef National Institute for Subatomic Physics, Amsterdam, Netherlands\\
% ...
$^{40}$Instituto Galego de F{\'\i}sica de Altas Enerx{\'\i}as (IGFAE), Universidade de Santiago de Compostela, Santiago de Compostela, Spain\\
% ...
$^{44}$Physik-Institut, Universit{\"a}t Z{\"u}rich, Z{\"u}rich, Switzerland\\
% ...
$^{48}$H.H. Wills Physics Laboratory, University of Bristol, Bristol, United Kingdom\\
% ...
$^{50}$Department of Physics, University of Warwick, Coventry, United Kingdom\\
% ...
$^{54}$Oliver Lodge Laboratory, University of Liverpool, Liverpool, United Kingdom\\
% ...
$^{62}$Syracuse University, Syracuse, NY, United States\\
% ...
$^{77}$University of Michigan, Ann Arbor, MI, United States, associated to $^{62}$\\
\bigskip
% ...
}
\end{flushleft}

\end{document}